\documentclass[%
 superscriptaddress,
 nofootinbib,
 amsmath,amssymb,
 aps,
twocolumn
]{revtex4-2}

\usepackage{graphicx}
\usepackage{bm}
\usepackage{xcolor}
\usepackage{amsmath}
\usepackage{kotex}
\usepackage{amsthm}
\usepackage{amsmath}
\usepackage{braket}
\usepackage[T1]{fontenc}

\newcommand{\Haf}{\mathrm{Haf}}
\newcommand{\Per}{\mathrm{Per}}
\newcommand{\lHaf}{\mathrm{lHaf}}

\begin{document}

\preprint{APS/123-QED}



\title{Combinatorial Matrix Functions in Ising Models \\and Quantum Circuit Representation}

\author{Minhyeok Kang}
\affiliation{SKKU Advanced Institute of Nanotechnology (SAINT), Sungkyunkwan University, Suwon 16419, South Korea}

\author{Gwonhak Lee}
\affiliation{IBM Quantum, Seoul 07335, Republic of Korea}

\author{Youngrong Lim}
\affiliation{Department of Physics, Chungbuk National University, Cheongju, Chungbuk 28644, Korea}

\author{Joonsuk Huh}
\email{joonsukhuh@gmail.com}
\affiliation{Department of Chemistry, Yonsei University, Seoul 03722, Republic of Korea}
\affiliation{Department of Quantum Information, Yonsei University, Incheon 21983, Republic of Korea}
\affiliation{Department of Computational Science and Engineering, Yonsei
University, Seoul 03722, Republic of Korea}




\date{\today}

\begin{abstract}
Permanents, hafnians, and loop-hafnians are combinatorial matrix functions defined on the perfect matchings of graphs. These matrix functions arise in the quantum amplitudes for output photon configurations in linear optical networks, which has led both to sampling hardness results and to a variety of applications. The same matrix functions also appear in Ising models, since Ising models naturally encode graphs, and the matrix functions emerge from quantum interference. However, previous results have been restricted to balanced bipartite graphs, and hence to the permanent. In this paper, we extend the graph structure of the Ising model so that all three matrix functions can be obtained. In particular, self-loops can be encoded by introducing additional spins, and the loop-hafnian is obtained from a superposition of spin states, which we call the loop-hafnian state. We show how to recover the hafnian and permanent cases as the number of self-loops decreases. Moreover, we show that the Ising dynamics we consider can be simulated on a quantum circuit with \(O(n^2)\) gates for a \(2n \times 2n\) real symmetric matrix. Specifically, we propose a quantum circuit that maps the loop-hafnian state to a basis state with \(O(n^2)\) gates, and show that, by an adaptive measurement, the loop-hafnian states for all subsets of vertices can be measured simultaneously.
\end{abstract}

\maketitle


\section{Introduction}

Permanents, hafnians, and loop-hafnians are combinatorial matrix functions defined as sums over the perfect matchings of a graph, with the matrix entries as edge weights, where the class of graphs determines the function: balanced bipartite graphs, simple graphs, and loop-augmented graphs, respectively. These matrix functions arise in the quantum amplitudes for output photon configurations in linear optical networks, with the specific function determined by the input state. This correspondence has made linear optical networks a central platform in two respects. On the one hand, since computing these matrix functions is \#P-hard~\cite{Valiant1979,bjorklund2019faster}, the sampling hardness of linear optical networks has been established, giving rise to boson sampling~\cite{aaronson2011computational} and Gaussian boson sampling~\cite{Hamilton2017,Kruse2019detailed,Li2025}. On the other hand, the properties of these matrix functions have enabled linear optical networks to be applied to a wide range of problems, such as molecular vibronic 
spectroscopy~\cite{huh2015,Huh2017,Clements2018,quesada2019franck,Zhu2024},  
molecular docking~\cite{Banchieaax1950,Yu2023}, 
and graph problems~\cite{Arrazola2018Dense,Schuld2020,Sempere2022Experimentally,Deng2023Solving}.
Moreover, the loop-hafnian arising in linear optical networks is related to the matching polynomial, which appears in 
monomer-dimer systems~\cite{Heilmann1972}.

The same matrix functions also appear in Ising models. The Ising model is naturally associated with a graph: spins correspond to vertices, couplings to edges, and coupling strengths to edge weights. \citet{Fefferman_permanent} showed that, for an Ising model of \(2n\) spins whose underlying graph is a balanced bipartite graph, the matrix element of the \(n\)th power of the Hamiltonian between the all spin-down and all spin-up states is proportional to the permanent of the coupling matrix. \citet{park2023hardness} further showed that several other quantum spin models on balanced bipartite graphs also give the permanent. However, existing results on quantum spin models have been restricted to balanced bipartite graphs, and hence to the permanent.

\begin{figure*}[htb!]
    \centering
    \includegraphics[width=1.0\linewidth]{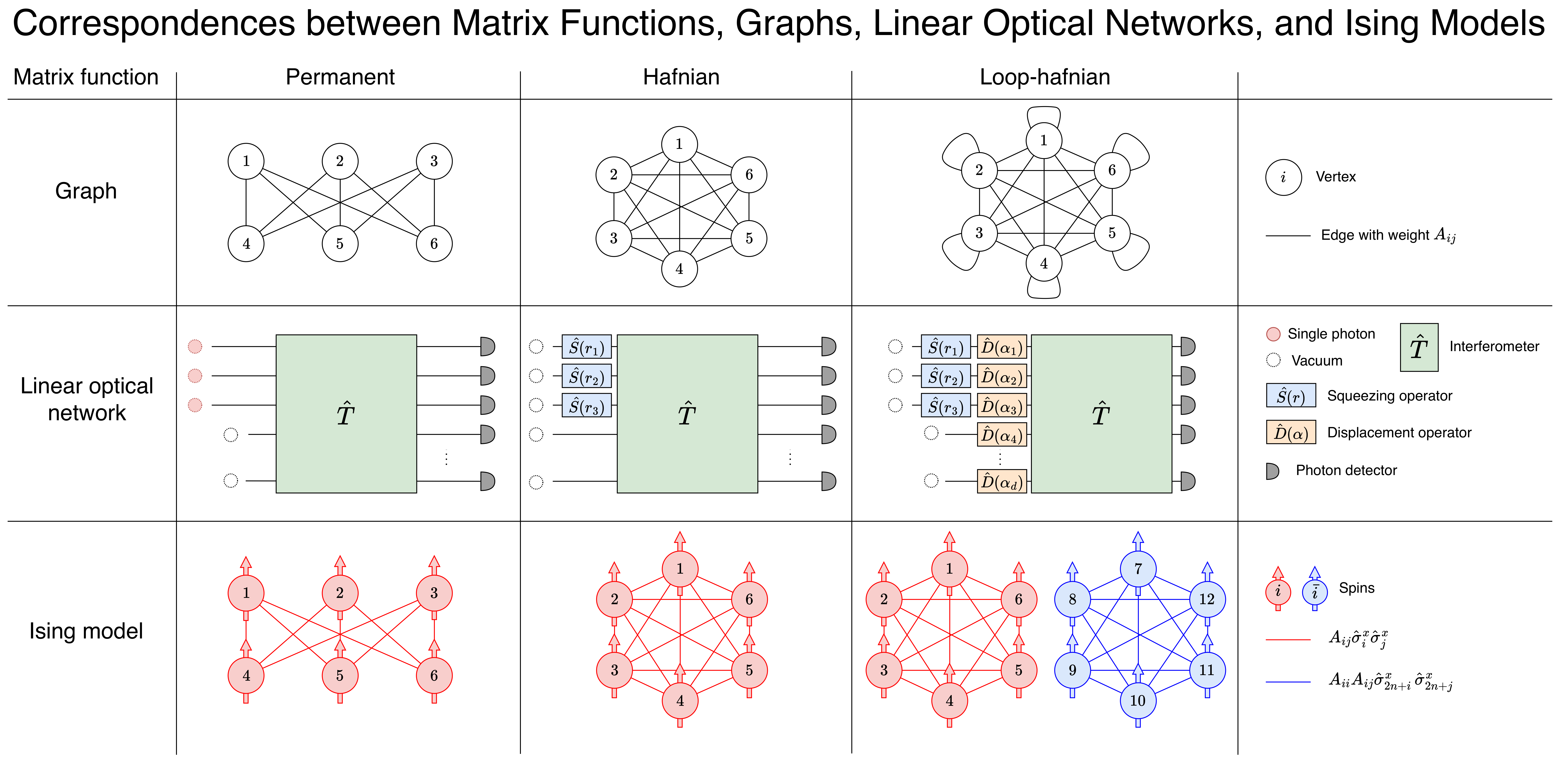}
    \caption{
    Graphical summary of the correspondences between matrix functions, graph structures, linear optical networks, and Ising models. Each matrix function is defined as a sum over the perfect matchings of a particular graph, with the matrix entries as edge weights. The matrix function depends on the input state in linear optical networks and on the interaction structure in Ising models. For the permanent and hafnian cases, the Ising model follows the graph exactly. In contrast, for the loop-hafnian case, the self-loops are encoded by introducing additional spins.}
    \label{fig:matrixandgraph}
\end{figure*}

In this work, we extend the Ising model beyond balanced bipartite graphs and show that all three combinatorial matrix functions can be represented as matrix elements of powers of the Hamiltonian. Not only the edges between different vertices but also self-loops can be encoded in the Ising model by introducing at most \(2n\) additional spins. The combinatorics underlying each matrix function emerges naturally from quantum interference. Fig.~\ref{fig:matrixandgraph} shows a graphical summary of the correspondences between matrix functions, graphs, linear optical networks, and Ising models. 

The construction for the loop-hafnian case naturally contains the hafnian and the permanent cases, since the classes of graphs corresponding to the three matrix functions are nested, and each class corresponds to a specific form of the matrix.
While the matrix function is obtained from a single spin state for the hafnian and the permanent cases, the loop-hafnian is obtained from a superposition of spin states, which we call the loop-hafnian state. We show that as the number of self-loops decreases, the loop-hafnian state becomes simpler and eventually reduces to a spin state, recovering the hafnian and the permanent cases.

Moreover, we show how to simulate the Ising dynamics on a quantum circuit, with each spin represented by a qubit and each spin state by a basis state. Since all terms in the Ising model mutually commute, the time evolution can be implemented on a quantum circuit by \(O(n^2)\) two-qubit gates without Trotter error. The remaining issue is the readout: for the hafnian and permanent cases it suffices to measure in the basis, whereas the loop-hafnian state is a superposition of basis states. We construct a measurement circuit that maps the loop-hafnian state to a basis state with \(O(p^2)\) gates, where \(p \leq 2n\) is the number of nonzero diagonal elements, and show that, by an adaptive measurement, the loop-hafnian states for all subsets of vertices can be measured simultaneously. Consequently, the Ising dynamics for all three matrix functions can be simulated and measured on a quantum circuit with \(O(n^2)\) gates.

This paper is organized as follows. Sec.~\ref{sec:setup} introduces the Ising model we consider and the notation used throughout this paper. Sec.~\ref{sec:zero_diagonal} treats real symmetric matrices with zero diagonal, which give the hafnian and permanent. Sec.~\ref{sec:non_zero_diagonal} treats general real symmetric matrices, which give the loop-hafnian. We also show how the general case reduces to the zero-diagonal case as the number of nonzero diagonal elements decreases. Sec.~\ref{sec:simulation} presents the simulation of the Ising dynamics on quantum circuits. Finally, Sec.~\ref{sec:conclusion} presents our conclusion and discusses future directions.

\section{Setup}
\label{sec:setup}
\begin{figure}
    \centering
    \includegraphics[width=1.0\linewidth]{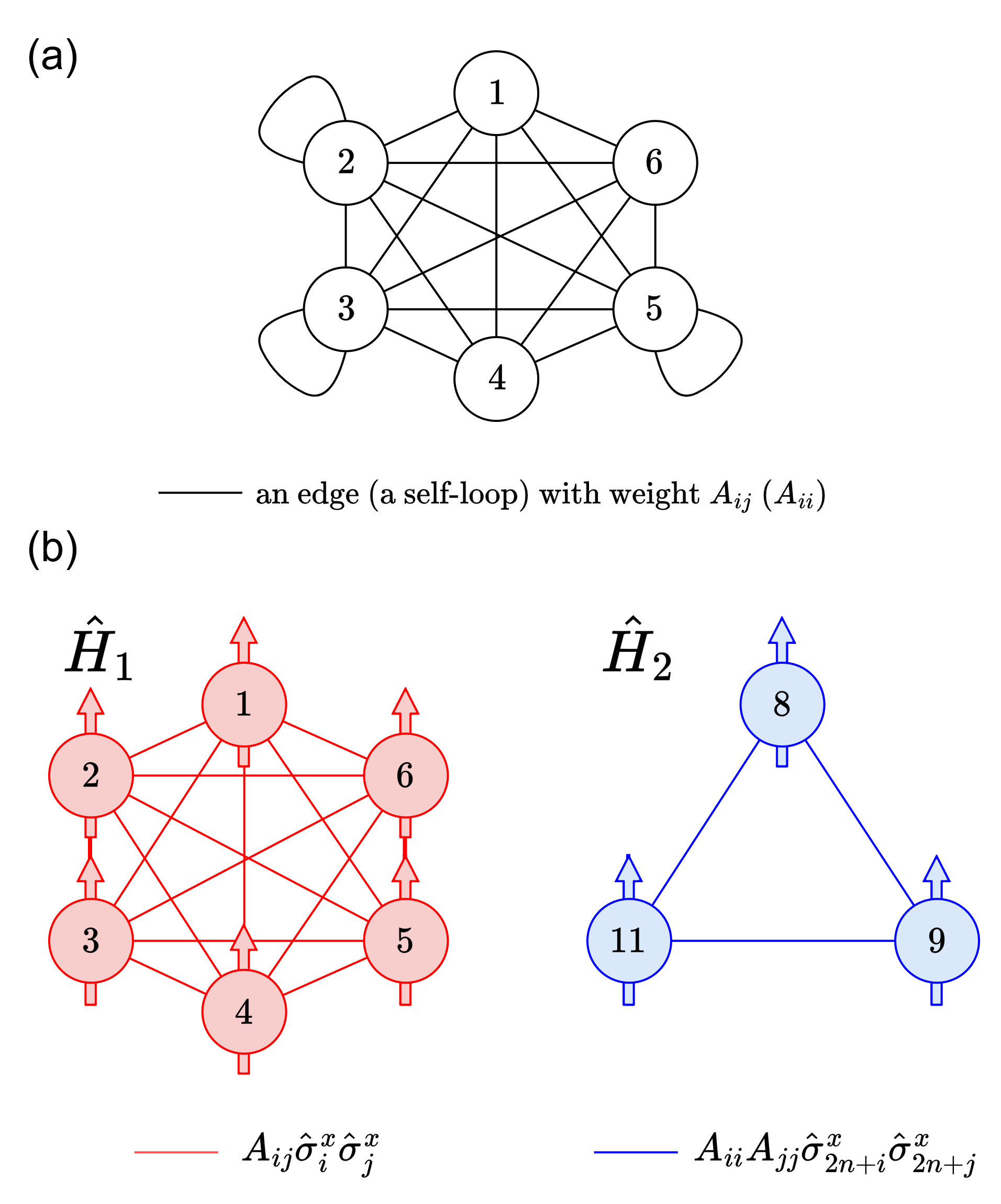}
    \caption{(a) A loop-augmented graph on \(2n = 6\) vertices with the adjacency matrix \(A\), where the vertices in \(L = \{2,3,5\}\) (\(p= 3\)) have self-loops. (b) The corresponding Ising model on \(2n+p = 9\) spins. \(\hat{H}_1\) (red) acts on spins \(1\)--\(6\) and encodes the edge weights, and \(\hat{H}_2\) (blue) acts on spins \(2n+i = 8,9,11\) for \(i \in L\) and encodes the products of self-loop weights.}
    \label{fig:model}
\end{figure}

Let \(A\) be a \(2n \times 2n\) real symmetric matrix, and let \(L \subseteq [2n]\) be the set of indices \(i\) with \(A_{ii} \neq 0\), with \(p := |L|\). We regard \(A\) as the adjacency matrix of a graph on \(2n\) vertices: the edge between vertices \(i\) and \(j\) has weight \(A_{ij}\), and each vertex \(i \in L\) has a self-loop of weight \(A_{ii}\). We consider the corresponding Ising model on \(2n+p\) spins, whose spin-up and spin-down states are denoted by \(\ket{\uparrow}\) and \(\ket{\downarrow}\), respectively. Instead of the conventional \(\hat{\sigma}^z\hat{\sigma}^z\), we take the two-body interaction to be \(\hat{\sigma}^x\hat{\sigma}^x\), so that each interaction term flips two spins. The last \(p\) spins are labeled by \(2n + i\) with \(i \in L\). The Hamiltonian is
\begin{align}
    \label{eqn: Hamiltonian}
    \hat{H} = \hat{H}_1 + \hat{H}_2
\end{align}
where
\begin{gather}
    \label{eqn: H_1}
    \hat{H}_1 = \sum_{1 \leq i < j \leq 2n}A_{ij}\hat{\sigma}_i^x\hat{\sigma}_j^x, \\
    \label{eqn: H_2}
    \hat{H}_2 = \sum_{\substack{i,j \in L \\ i <j} }A_{ii}A_{jj}\hat{\sigma}_{2n + i}^x\hat{\sigma}_{2n+j}^x.
\end{gather}
\(\hat{H}_1\) acts on the first \(2n\) spins and encodes the edge weights. \(\hat{H}_2\) acts on the last \(p\) spins and encodes the products of pairs of self-loop weights (see Fig.~\ref{fig:model}). Since the two parts act on disjoint sets of spins, they commute.

Note that when at most one diagonal element is nonzero (\(p \leq 1\)), \(\hat{H}_2\) contains no term and the system reduces to a \(2n\)-spin system governed by \(\hat{H}_1\). Since each term in \(\hat{H}\) is of the form \(\hat{\sigma}_i^x\hat{\sigma}_j^x\) with \(i\neq j\), each term in \(\hat{H}^k\) is a product of at most \(2k\) \(\hat{\sigma}^x\) operators and thus flips at most \(2k\) spins.

Throughout this paper, we use the following notation. For a \(k\)-spin system, we define the state \(\ket{S}\) for a subset \(S \subseteq [k]\) as the spin state in which the spins indexed by \(S\) are down and the other spins are up. For the (\(2n +p\))-spin system introduced above, we write \(\ket{S,T} = \ket{S}\ket{T}\) for \(S \subseteq [2n]\) and \(T \subseteq L\), where \(\ket{S}\) is the spin state of the first \(2n\) spins and \(\ket{T}\) denotes the spin state of the last \(p\) spins in which the spins \(2n + i\) with \(i \in T\) are down.

Given a real matrix \(M\) of size \(k \times k\), we denote by \(M_{S,T}\) the submatrix of \(M\) containing the rows in \(S\) and the columns in \(T\). Also, we denote by \(M_S\) the principal submatrix of \(M\) induced by \(S\), i.e., \(M _S = M_{S,S}\). Finally, we adopt the convention that products of operators are ordered from right to left: \(\prod_{i=1}^{t}\hat{O}_i = \hat{O}_t\cdots\hat{O}_2\hat{O}_1\), and we define \((-1)!! := 1\).

\section{Zero diagonal: permanents and hafnians}
\label{sec:zero_diagonal}

We first consider a \(2n \times 2n\) real symmetric matrix \(A\) with zero diagonal. In this case, the Hamiltonian is \(\hat{H} = \hat{H}_1\).

Let \(S \subseteq [2n]\) with \(|S| = 2m\). Since \(\hat{H}^k\) flips at most \(2k\) spins, \(\braket{S|\hat{H}^k|\uparrow^{2n}}\) vanishes for all \(k < m\), and
\begin{align}
    \label{eqn:hafnian_result}
    \braket{S|\hat{H}^m|\uparrow^{2n}} = m!\Haf(A_S),
\end{align}
where \(\Haf(A_S)\) is the hafnian of \(A_S\) (see Appendix~\ref{app:appendix_A}). The hafnian of a \(2l \times 2l\) symmetric matrix \(M\) is defined by
\begin{align}
    \Haf(M) = \sum_{\rho \in \mathrm{PMP}_{2l}}\prod_{\{i,j\} \in \rho} M_{ij},
\end{align}
where \(\mathrm{PMP}_{2l}\) is the set of perfect matchings of \([2l]\), e.g., \(\{\{1,2\},\{3,4\},\{5,6\}\} \in \mathrm{PMP}_6\). For a simple graph on \(2l\) vertices with adjacency matrix \(M\), \(\Haf(M)\) is the sum over all perfect matchings of the product of the edge weights. Therefore, \(\braket{S|\hat{H}^m|\uparrow^{2n}}\) is proportional to the hafnian of the adjacency matrix of the subgraph induced by \(S\).

For a balanced bipartite graph on \(2n\) vertices, the adjacency matrix has the form 
\begin{align}
\label{eqn:block_mat}
    A = \begin{pmatrix}
        O_n & B \\ B^\top & O_n
    \end{pmatrix},
\end{align}
where \(O_n\) is the \(n \times n\) zero matrix and \(B\) is an \(n \times n\) real matrix, called the biadjacency matrix of the graph. The Hamiltonian becomes
\begin{align}
    \hat{H} = \sum_{i,j=1}^{n}B_{ij}\hat{\sigma}_i^x\hat{\sigma}_{n+j}^x,
\end{align}
which is identical to the model proposed in Refs.~\cite {Fefferman_permanent,huh2025quantum,park2023hardness}.

In this case, for two subsets \(T,V \subseteq [n]\) with \(|T|=|V|=m\), the matrix element of \(\hat{H}^m\) between \(\ket{T\cup (n+V)}\) and \(\ket{\uparrow^{2n}}\) is
\begin{align}
    \braket{T\cup (n+V)|\hat{H}^m|\uparrow^{2n}} = m!\Per(B_{T,V}),
\end{align}
where \(\Per(B_{T,V})\) is the permanent of \(B_{T,V}\). This follows from Eq.~\eqref{eqn:hafnian_result}, since \(\Haf(A_{T \cup (n+V)}) = \Per(B_{T,V})\) for \(A\) of the form of Eq.~\eqref{eqn:block_mat}. The permanent of a \(k \times k\) matrix \(M\) is
\begin{align}
    \Per(M) = \sum_{\sigma \in S_k}\prod_{i=1}^kM_{i\sigma(i)},
\end{align}
where \(S_k\) is the symmetric group on \([k]\). For a balanced bipartite graph on \(2k\) vertices with biadjacency matrix \(M\), \(\Per(M)\) is the sum over all perfect matchings of the product of the edge weights. Therefore, \(\braket{T\cup (n+V)|\hat{H}^m|\uparrow^{2n}}\) is proportional to the permanent of the biadjacency matrix of the subgraph induced by \(T \cup (n+V)\).

\section{Nonzero diagonal: loop-hafnians}
\label{sec:non_zero_diagonal}

We now consider a general \(2n \times 2n\) real symmetric matrix \(A\), and let \(L \subseteq[2n]\) be the set of indices \(i\) with \(A_{ii} \neq 0\),  with \(|L| = p\). In this case, \(\hat{H}_2\) must be included, and the full Hamiltonian is given by Eq.~\eqref{eqn: Hamiltonian} acting on the (\(2n+p\))-spin system. The loop-hafnian of a \(k \times k\) symmetric matrix \(M\) is defined by
\begin{align}
    \lHaf(M) = \sum_{i=0}^{k}\sum_{\substack{S \subseteq [k] \\ |S| = i}}\left(\prod_{j \in [k]\setminus S}M_{jj}\right)\Haf(M_S),
\end{align}
where we set \(\Haf(M_{\emptyset}) := 1\). Note that, by definition of the hafnian, only subsets \(S \subseteq [k]\) with an even number of elements contribute. For a loop-augmented graph on \(k\) vertices with adjacency matrix \(M\), \(\lHaf(M)\) is the sum over all perfect matchings, in which self-loops are allowed, of the product of the edge and self-loop weights. The size \(k\) does not need to be an even number, since a vertex can be matched to itself through its self-loop. Nevertheless, the same loop-hafnian can be obtained by adding a dummy vertex with only a self-loop of weight \(1\), whose adjacency matrix is
\begin{align}
    M' = \begin{pmatrix}
        M & \vec{0} \\ \vec{0}^\top & 1
    \end{pmatrix},
\end{align}
where \(\vec{0}\) is the \(k\)-dimensional zero vector, so that \(\lHaf(M') = \lHaf(M)\). Hence, without loss of generality, we assume \(k = 2l\) is even, so that
\begin{align}
    \lHaf(M) = \sum_{i=0}^l \sum_{\substack{S \subseteq [2l] \\ |S| = 2i}}\left(\prod_{j \in [2l]\setminus S} M_{jj}\right)\Haf(M_S).
\end{align}

Let \(S \subseteq [2n]\) with \(|S| =2m\), and choose a subset \(V \subseteq S\cap L\) with \(|V|= 2k\). Then \(\braket{S\setminus V,V|\hat{H}^m|\uparrow^{2n+p}}\) is given by (see Appendix~\ref{app:appendix_B})
\begin{multline}
\label{eqn:weighted_hafnian}
    \braket{S\setminus V,V|\hat{H}^m|\uparrow^{2n+p}} \\= m!(2k-1)!!\left(\prod_{i \in  V}A_{ii}\right)\Haf(A_{S \setminus V}).
\end{multline}
Note that, since \(\ket{S\setminus V,V}\) has \(2m\) down spins, \(\braket{S\setminus V,V|\hat{H}^q|\uparrow^{2n+p}} = 0\) for \(q < m\).

Eq.~\eqref{eqn:weighted_hafnian} shows that each term of the loop-hafnian of \(A_S\) appears in the matrix element for a different spin state. The subset \(V\) specifies which vertices of \(S\) are matched to themselves, and the remaining vertices in \(S \setminus V\) are matched by edges. The loop-hafnian itself is therefore obtained not from a single spin state but from a superposition of them, in which the weight \(1/(2k-1)!!\) compensates for the factor \((2k-1)!!\) in Eq.~\eqref{eqn:weighted_hafnian}. We define the loop-hafnian state \(\ket{\psi_{S,L}}\) as
\begin{align}
    \label{eqn:loop_hafnian_state}
    \ket{\psi_{S,L}} := \frac{1}{N_{S,L}}\sum_{k=0}^{\lfloor|S\cap L|/2\rfloor}\frac{1}{(2k-1)!!}\sum_{\substack{V \subseteq S\cap L \\ |V| = 2k}}\ket{S\setminus V,V}
\end{align}
with the normalization factor \(N_{S,L}\)
\begin{align}
\label{eqn:normalization}
    N_{S,L}^2 = \sum_{k=0}^{\lfloor|S\cap L|/2\rfloor} \frac{1}{(2k-1)!!^2}\binom{|S\cap L|}{2k}.
\end{align}

As a result,
\begin{align}
\label{eqn:loop_hafnian_result}
    \braket{\psi_{S,L}|\hat{H}^m|\uparrow^{2n+p}} = \frac{m!}{N_{S,L}}\lHaf(A_S).
\end{align}
The derivations of Eqs.~\eqref{eqn:normalization} and \eqref{eqn:loop_hafnian_result} are given in Appendix~\ref{app:appendix_C}.

As \(|S\cap L|\) decreases, the number of spin states in \(\ket{\psi_{S,L}}\) decreases. For \(S \cap L = \emptyset\), we have \(\ket{\psi_{S,L}} = \ket{S}\ket{\uparrow^{p}}\) and \(\braket{\psi_{S,L}|\hat{H}^m|\uparrow^{2n+p}} = m!\Haf(A_S)\). In particular, when at most one diagonal element is nonzero (\(p\leq1\)), \(\hat{H}_2\) vanishes, the system reduces to the \(2n\)-spin system, and \(\ket{\psi_{S,L}}= \ket{S}\), which recovers the hafnian case of Sec.~\ref{sec:zero_diagonal}. Moreover, if \(A\) has the block form of Eq.~\eqref{eqn:block_mat}, then for \(S = T \cup (n+V)\) with \(T,V \subseteq [n]\) and \(|T|=|V|=m\), we obtain \(\braket{\psi_{S,L}|\hat{H}^m|\uparrow^{2n}} = m!\Per(B_{T,V})\), which recovers the permanent case.

\section{Simulation of Ising dynamics on quantum circuits}
\label{sec:simulation}

The Ising dynamics we consider can be simulated on a quantum circuit with \({O}(n^2)\) gates. We represent each spin by a qubit, with \(\ket{0}\) and \(\ket{1}\) corresponding to \(\ket{\uparrow}\) and \(\ket{\downarrow}\), respectively, so that the Pauli operator \(\hat{\sigma}_j^x\) is represented by the X gate \(X_j\) and the state \(\ket{S}\) is the basis state in which the qubits in \(S\) are \(\ket{1}\). Although \(\hat{H}^m\) cannot in general be implemented directly on a quantum circuit~\cite{Aulicino2022}, the transition amplitude after a propagation time \(t\) provides an approximation to the matrix element of \(\hat{H}^m\):
\begin{multline}
\label{eqn:trans_amp}
    \braket{\psi_{S,L}|e^{-i\hat{H}t}|0^{2n+p}} = \sum_{a=0}^{\infty}\frac{(-it)^a}{a!}\braket{\psi_{S,L}|\hat{H}^a|0^{2n+p}}
    \\= \sum_{a=0}^{\infty}\frac{(-it)^{m+a}}{(m+a)!}\braket{\psi_{S,L}|\hat{H}^{m+a}|0^{2n+p}}
    \\= \frac{(-it)^{m}}{N_{S,L}}\lHaf(A_S) + \sum_{a=1}^{\infty}\frac{(-it)^{m+a}}{(m+a)!}\braket{\psi_{S,L}|\hat{H}^{m+a}|0^{2n+p}}
    \\=\frac{(-it)^m}{N_{S,L}}\left[\lHaf(A_S) + O(t)\right].
\end{multline}
Hence, for short \(t\), the leading contribution to the transition amplitude is proportional to \(\lHaf(A_S)\) up to an additive error of \(O(t)\) arising from the higher-order terms. Its transition probability thus becomes
\begin{multline}
    \label{eqn:trans_prob}
    p_{S,L}(t):=|\braket{\psi_{S,L}|e^{-i\hat{H}t}|0^{2n + p}}|^2 
    \\=\frac{t^{2m}}{N_{S,L}^2}\left[\lHaf(A_S)^2 + O(t^2)\right],
\end{multline}
which is proportional to the squared loop-hafnian up to an additive error of \(O(t^2)\).  For the balanced bipartite case, the sampling hardness of this dynamics has been studied in Refs.~\cite{Fefferman_permanent,park2023hardness}.

Since all terms of \(\hat{H}\) are of the form \(\hat{\sigma}^x\hat{\sigma}^x\) and mutually commute, the time evolution operator \(e^{-i\hat{H}t}\) can be implemented by a product of RXX gates \(\exp(-i\theta X_jX_k/2)\) with \(O(n^2)\) gates and without Trotter error.

In the zero-diagonal case, the matrix functions appear in matrix elements between basis states, so it suffices to measure in the basis. In contrast, the loop-hafnian state is a superposition of basis states, and we need a measurement circuit that maps it to a basis state.

In Appendix~\ref{app:appendix_D} we show that the loop-hafnian state \(\ket{\psi_{S,L}}\) can be mapped to the basis state \(\ket{S}\ket{0^p}\) with \(O(p^2)\) gates. Applying the \(p\) CNOT gates \(CX^{(2n+i,i)}\), with qubit \(2n+i\) as control and qubit \(i\) as target, for all \(i \in L\), separates \(\ket{\psi_{S,L}}\) into \(\ket{S}\) on the first \(2n\) qubits and a state \(\ket{\phi_{S,L}}\) on the last \(p\) qubits. The state \(\ket{\phi_{S,L}}\) is then mapped to \(\ket{0^p}\) by an operator \(\hat{G}_{S,L}\), which can be implemented with \(O(p^2)\) gates.

\begin{figure}
    \centering
    \includegraphics[width=1.0\linewidth]{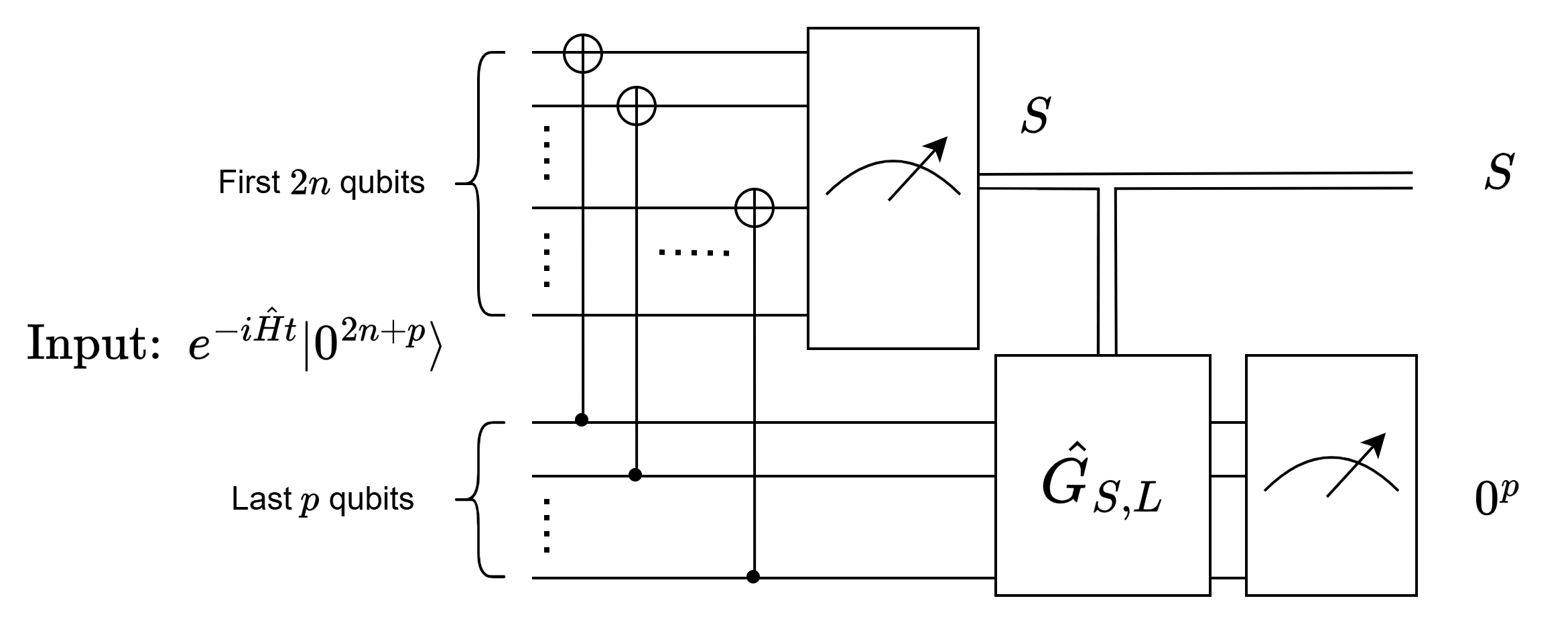}
    \caption{The measurement circuit for the loop-hafnian states \(\ket{\psi_{S,L}}\) for all \(S \subseteq [2n]\). After the \(p\) CNOT gates, the first \(2n\) qubits are measured, and the operator \(\hat{G}_{S,L}\) determined by the outcome \(S\) is applied to the last \(p\) qubits before they are measured. \(\hat{G}_{S,L}\) can be implemented with \(O(p^2)\) gates, as detailed in Appendix~\ref{app:appendix_D}, so that all loop-hafnian states can be measured with \(O(p^2)\) gates.}
    \label{fig:measure}
\end{figure}

Although the operator \(\hat{G}_{S,L}\) depends on \(S\), the CNOT gates do not. Therefore, we can perform an adaptive measurement: after the \(p\) CNOT gates, we measure the first \(2n\) qubits, obtain an outcome \(S\), then apply \(\hat{G}_{S,L}\) to the last \(p\) qubits according to that outcome. In this way, the probability of obtaining the outcome \(\ket{S}\ket{0^p}\) equals \(|\braket{\psi_{S,L}|e^{-i\hat{H}t}|0^{2n+p}}|^2\) for every \(S\) simultaneously. Fig.~\ref{fig:measure} shows the resulting circuit. Consequently, the Ising dynamics, including the measurement of the loop-hafnian states, can be simulated with \(O(n^2)\) gates.

The qubit connectivity required for the simulation depends on the structure of \(A\). A nonzero off-diagonal element \(A_{ij}\) requires a connection between qubits \(i\) and \(j\), and a pair of nonzero diagonal elements \(A_{ii}\) and \(A_{jj}\) requires a connection between qubits \(2n+i\) and \(2n+j\). In addition, the measurement circuit requires a connection between qubits \(i\) and \(2n+i\) for each \(i \in L\). Thus, for the permanent, hafnian, and loop-hafnian, the connectivity required is at most a complete balanced bipartite graph \(K(n,n)\), a complete graph on \(2n\) vertices, and the union of a complete graph on \(2n\) vertices and a complete graph on \(p\) vertices joined by the \(p\) edges \((i,2n+i)\), respectively. A trapped-ion system is a suitable platform for this simulation, since any pair of qubits admits a direct two-qubit gate~\cite{Blatt2012,monroe2021programmable,Wright2019,Debnath2016,kang2025}.

\section{Conclusion and Discussion}
\label{sec:conclusion}

We have shown how the Ising model gives all three combinatorial matrix functions. When the matrix entries are encoded as coupling strengths and the output states are chosen appropriately, the matrix functions appear as matrix elements of powers of the Hamiltonian. Moreover, we have shown that the corresponding Ising dynamics can be implemented and read out on a quantum circuit with \(O(n)\) qubits and \(O(n^2)\) gates for a real symmetric matrix of size \(2n \times 2n\).

Let us consider the difference between linear optical networks and Ising models from the matrix function perspective. In linear optical networks, the matrix functions appear in the transition amplitudes of photon configurations, and the matrices are complex-valued since they are determined by the unitary matrix of the interferometer~\cite{aaronson2011computational,Hamilton2017,Kruse2019detailed}. Therefore, the transition probability is proportional to the squared modulus of matrix functions, with a factor that depends on the photon distribution. When every output mode contains at most one photon, the probability is given by the squared modulus up to a normalization factor independent of the output pattern. This direct relation between the output distribution and the matrix functions underlies the sampling problems and applications mentioned in the introduction.

In contrast, Ising models are restricted to real symmetric matrices, since the matrix entries are encoded as coupling strengths of the Hamiltonian. Moreover, the matrix functions appear in matrix elements of powers of the Hamiltonian, which cannot be obtained from the Ising dynamics directly. As shown in Eq.~\eqref{eqn:trans_prob}, the transition probability of a state with \(2m\) down spins is proportional to \(t^{2m}\) times the squared matrix function, up to an additive error of \(O(t^2)\). Therefore, there is a trade-off between the error and the probability scale: as the propagation time decreases, the error decreases, but so does the probability itself, and vice versa. Consequently, the applications proposed for linear optical networks cannot be transferred directly to Ising models.

Nevertheless, the sampling hardness arguments for linear optical networks may be transferred to Ising models from the matrix function perspective. Boson sampling~\cite{aaronson2011computational} and Gaussian boson sampling~\cite{Hamilton2017,Kruse2019detailed} consider the approximate sampling hardness, in which an additive error on the output probabilities translates into an additive error on the squared modulus of the matrix functions. The hardness is established in the average case over Gaussian matrices, under conjectures on the matrix functions of Gaussian matrices. Since the transition probability in our model is proportional to the squared matrix function up to an additive error, the sampling hardness of Ising models could be discussed by applying the approximate sampling hardness arguments of boson sampling and Gaussian boson sampling. The sampling hardness of Ising models has previously been treated in the framework of instantaneous quantum polynomial-time (IQP) circuits~\cite{bremner2011classical,bremner2016average}, where the hardness is derived from the \#P-hardness of approximating the imaginary-temperature Ising partition function~\cite{Fujii2017,Goldberg2017}; the matrix function perspective offers a different route to the sampling hardness of Ising models. While \citet{park2023hardness} have studied the permanent case in the average-case setting, extending the argument to the general Ising model is left for future work.

\begin{acknowledgments}
This work was partly supported by the following multiple funding sources:
(i) Basic Science Research Program through the National Research Foundation of Korea (NRF), funded by the Ministry of Science and ICT (RS-2023-NR068116, RS-2025-03532992). (ii) Institute for Information \& Communications Technology Promotion (IITP) grant funded by the Korea government (MSIP) (No. 2019-0-00003, No. RS-2024-00437284), which focuses on the research and development of core technologies for programming, running, implementing, and validating fault-tolerant quantum computing systems. (iii) Yonsei University Research Fund under project number 2025-22-0140. (iv) The Korean ARPA-H Project via the Korea Health Industry Development Institute (KHIDI); Ministry of Health and Welfare, Republic of Korea (RS-2025-25456722).
\end{acknowledgments}

\newpage
\onecolumngrid

\section*{Appendix}
\appendix

\section{Derivation of Eq.~\eqref{eqn:hafnian_result}}
\label{app:appendix_A}
In this section, we show that \(\braket{S|\hat{H}^{m}|\uparrow^{2n}} = m!\Haf(A_S)\) for all \(m \leq n\) and for all \(S \subseteq [2n]\) with \(|S| = 2m\) using mathematical induction.

For \(m = 1\), without loss of generality, assume that \(S = \{1,2\}\). Then
\begin{align}
    \braket{S|\hat{H}|\uparrow^{2n}} = \sum_{1 \leq i < j \leq 2n}A_{ij}\braket{S|\hat{\sigma}_i^x\hat{\sigma}_j^x|\uparrow^{2n}} = \sum_{1 \leq i < j \leq 2n}A_{ij}\braket{S|\{i,j\}}
     = A_{12} = \Haf(A_S).
\end{align}

Assume that \(\braket{S|\hat{H}^m|\uparrow^{2n}} = m!\Haf(A_S)\) for all subsets \(S \subseteq [2n]\) with \(|S| = 2m\) for some \(m < n\). Let \(S' \subseteq [2n]\) with \(|S'| = 2m+2\). The hafnian of \(A_{S'}\) can be written as a sum of hafnians of the submatrices \(A_{S'\setminus \{i,j\}}\) for all \(i,j \in S'\):

\begin{align}
    \Haf(A_{S'}) = \frac{1}{m+1}\sum_{\{i,j\} \subseteq S'}A_{ij}\Haf(A_{S'\setminus \{i,j\}}).
\end{align}
The factor \(\frac{1}{m+1}\) compensates for the fact that each perfect matching of \(S'\) is counted once for each of its \(m+1\) edges. 

\(\braket{S'|\hat{H}^{m+1}|\uparrow^{2n}}\) is 
\begin{align}
    \braket{S'|\hat{H}^{m+1}|\uparrow^{2n}} = \braket{S'|\hat{H}\hat{H}^m|\uparrow^{2n}}
    = \sum_{1 \leq i <j \leq 2n}A_{ij}\braket{S'|\hat{\sigma}_i^x\hat{\sigma}_j^x\hat{H}^m|\uparrow^{2n}}.
\end{align}
Since \(\ket{S'}\) has \(2m+2\) down spins, \(\hat{\sigma}_i^x\hat{\sigma}_j^x\ket{S'}\) has \(2m\) (\(i,j \in S'\)), \(2m+4\) (\(i,j \notin S'\)), or \(2m+2\) down spins (otherwise). Since \(\hat{H}^m\ket{\uparrow^{2n}}\) contains at most \(2m\) down spins, only the terms with \(i,j \in S'\) survive:
\begin{multline}
    \sum_{1 \leq i <j \leq 2n}A_{ij}\braket{S'|\hat{\sigma}_i^x\hat{\sigma}_j^x\hat{H}^m|\uparrow^{2n}}
    = \sum_{\{i,j\} \subseteq S'}A_{ij}\braket{S'\setminus \{i,j\}|\hat{H}^m|\uparrow^{2n}}
    \\ = m!\sum_{\{i,j\} \subseteq S'}A_{ij}\Haf(A_{S'\setminus \{i,j\}}) = (m+1)!\Haf(A_{S'}).
\end{multline}
 
Therefore, \(\braket{S|\hat{H}^m|\uparrow^{2n}} = m!\Haf(A_S)\) for all \(m \leq n\) and for all subsets \(S \subseteq [2n]\) with \(|S| = 2m\).

\section{Derivation of Eq.~\eqref{eqn:weighted_hafnian}}
\label{app:appendix_B}

Consider the Hamiltonian Eq.~\eqref{eqn: Hamiltonian} of \(2n+p\) spins, where \(A_{ii}\neq 0\) for \(i \in L \subseteq [2n]\) with \(|L| = p\). Let \(S \subseteq [2n]\) with \(|S| = 2m\), and choose \(V \subseteq S\cap L\) with \(|V| = 2k\). Since \(\hat{H}_1\) and \(\hat{H}_2\) commute,
\begin{multline}
    \braket{S\setminus V,V|\hat{H}^m|\uparrow^{2n+p}} = \braket{S\setminus V,V|\left(\hat{H}_1+\hat{H}_2\right)^m|\uparrow^{2n+p}}
    =\sum_{j=0}^{m}\binom{m}{j}\braket{S\setminus V|\hat{H}_1^j|\uparrow^{2n}}\braket{V|\hat{H}_2^{m-j}|\uparrow^{p}}
    \\= \binom{m}{m-k}\braket{S\setminus V|\hat{H}_1^{m-k}|\uparrow^{2n}}\braket{V|\hat{H}_2^k|\uparrow^{p}},
\end{multline}
because \(|S\setminus V| = 2(m-k)\) and \(|V| = 2k\). Then \(\braket{S\setminus V|\hat{H}_1^{m-k}|\uparrow^{2n}} = (m-k)!\Haf(A_{S\setminus V})\) by Appendix~\ref{app:appendix_A}.  \(\hat{H}_2\) has the same form as \(\hat{H}_1\) with \(A\) replaced by \(A'\) where \(A'_{ij}= A_{ii}A_{jj}\) for \(i,j \in L\) and \(i \neq j\), and thus \(\braket{V | \hat{H}_2^{k}|\uparrow^{p}} = k!\Haf(A'_{V})\). Since every perfect matching of \(V\) contributes the same product \(\prod_{i \in V}A_{ii}\),
\begin{align}
    \braket{V|\hat{H}_2^{k}|\uparrow^{p}} = k!(2k-1)!!\prod_{i \in V}A_{ii},
\end{align}
where \((2k-1)!!\) is the number of perfect matchings of \(V\). As a result,
\begin{multline}
    \braket{S\setminus V,V|\hat{H}^m|\uparrow^{2n+p}} = \binom{m}{m-k}(m-k)!\Haf(A_{S\setminus V})k!(2k-1)!!\left(\prod_{i\in V}A_{ii}\right)
    \\= m! (2k-1)!!\left(\prod_{i \in V}A_{ii}\right)\Haf(A_{S\setminus V}).
\end{multline}

\section{Derivations of Eqs.~\eqref{eqn:normalization} and \eqref{eqn:loop_hafnian_result}}
\label{app:appendix_C}

In Sec.~\ref{sec:non_zero_diagonal}, we defined the loop-hafnian state \(\ket{\psi_{S,L}}\) for a subset \(S \subseteq [2n]\) with \(|S| = 2m\), where \(L \subseteq [2n]\) is the set of indices with \(A_{ii}\neq 0\) and \(|L|= p\). In this section, we derive Eqs.~\eqref{eqn:normalization} and \eqref{eqn:loop_hafnian_result}. Recall the loop-hafnian state \(\ket{\psi_{S,L}}\):

\begin{align}
\tag{\ref{eqn:loop_hafnian_state}}
    \ket{\psi_{S,L}} = \frac{1}{N_{S,L}}\sum_{k=0}^{\lfloor|S \cap L|/2\rfloor}\frac{1}{(2k-1)!!} \sum_{\substack{V \subseteq S\cap L \\ |V| = 2k}}\ket{S \setminus V,V}.
\end{align}

From Appendix~\ref{app:appendix_B}, \(\braket{\psi_{S,L}|\hat{H}^m|\uparrow^{2n+p}}\) becomes
\begin{multline}
    \braket{\psi_{S,L}|\hat{H}^m|\uparrow^{2n+p}} = \frac{1}{N_{S,L}}\sum_{k=0}^{\lfloor|S \cap L|/2\rfloor}\frac{1}{(2k-1)!!}\sum_{\substack{V \subseteq S\cap L \\ |V| = 2k}}\braket{S\setminus V,V|\hat{H}^m |\uparrow^{2n+p}}
    \\=\frac{1}{N_{S,L}}\sum_{k=0}^{\lfloor|S \cap L|/2\rfloor}\frac{m!(2k-1)!!}{(2k-1)!!}\sum_{\substack{V\subseteq S\cap L \\ |V| = 2k}}\left(\prod_{i \in V}A_{ii}\right)\Haf(A_{S\setminus V})
    =\frac{m!}{N_{S,L}}\sum_{k=0}^{\lfloor|S \cap L|/2\rfloor}\sum_{\substack{V \subseteq S\cap L \\ |V| = 2k}}\left(\prod_{i \in V}A_{ii}\right)\Haf(A_{S\setminus V}) 
    \\= \frac{m!}{N_{S,L}}\lHaf(A_S).
\end{multline}

The normalization factor \(N_{S,L}\) is determined by \(\braket{\psi_{S,L}|\psi_{S,L}} = 1\):
\begin{multline}
    \braket{\psi_{S,L}|\psi_{S,L}} = 
    \frac{1}{N_{S,L}^2}\sum_{k,k'=0}^{\lfloor|S \cap L|/2\rfloor}\frac{1}{(2k-1)!!}\frac{1}{(2k'-1)!!}\sum_{\substack{V,V' \subseteq S\cap L \\ |V| = 2k ,\;|V'|=2k'}}\braket{S\setminus V,V|S \setminus V',V'}
    \\=\frac{1}{N_{S,L}^2}\sum_{k=0}^{\lfloor|S \cap L|/2\rfloor}\frac{1}{(2k-1)!!^2}\sum_{\substack{V \subseteq S\cap L \\ |V| = 2k }}1
    = \frac{1}{N_{S,L}^2}\sum_{k=0}^{\lfloor|S \cap L|/2\rfloor}\frac{1}{(2k-1)!!^2} \binom{|S\cap L|}{2k} =1.
\end{multline}

Therefore,
\begin{align}
    N_{S,L}^2 = \sum_{k=0}^{\lfloor|S \cap L|/2\rfloor}\frac{1}{(2k-1)!!^2}\binom{|S\cap L|}{2k}.
\end{align}

\section{Measurement circuit for the loop-hafnian state}
\label{app:appendix_D}
In this section, we construct a circuit that maps the loop-hafnian state \(\ket{\psi_{S,L}}\) to a basis state. Applying \(p\) CNOT gates \(CX^{(2n+i,i)}\) for all \(i \in L\) to \(\ket{\psi_{S,L}}\) disentangles the first \(2n\) qubits from the last \(p\) qubits:
\begin{align}
    \prod_{i \in L}CX^{(2n+i,i)} \ket{\psi_{S,L}}
    =
    \frac{1}{N_{S,L}}\sum_{k=0}^{\lfloor|S \cap L|/2\rfloor}\frac{1}{(2k-1)!!}\sum_{\substack{ V \subseteq S\cap L \\ |V| = 2k}}\ket{S,V}
    =: \ket{S}\ket{\phi_{S,L}},
\end{align}
where 
\begin{align}
    \ket{\phi_{S,L}} = \frac{1}{N_{S,L}}\sum_{k=0}^{\lfloor|S \cap L|/2\rfloor}\frac{1}{(2k-1)!!} \sum_{\substack{V  \subseteq S\cap L \\ |V| = 2k}}\ket{V}.
\end{align}
If \(|S \cap L| \leq 1\), then \(\ket{\phi_{S,L}}=\ket{0^{p}}\) and the resulting state \(\ket{S}\ket{0^p}\) is already a basis state. We therefore assume \(|S \cap L| \geq 2\) in what follows.

For \(V \subseteq S\cap L\), the basis state \(\ket{V}\) of the last \(p\) qubits factorizes as \(\ket{V} = \ket{V}_{S\cap L}\ket{0^{p-|S\cap L|}}_{L\setminus S}\), where the subscripts indicate the qubits in \(S \cap L\) and \(L\setminus S\), respectively. Therefore, we can write \(\ket{\phi_{S,L}} = \ket{\phi_{|S\cap L|}}_{S\cap L}\ket{0^{p - |S\cap L|}}_{L\setminus S}\) with
\begin{align}
    \ket{\phi_{|S\cap L|}} = \frac{1}{N_{S,L}}\sum_{k=0}^{\lfloor |S \cap L|/2\rfloor}\frac{1}{(2k-1)!!}\sum_{\substack{V \subseteq [|S\cap L|]\\|V| = 2k}}\ket{V}.
\end{align}
For fixed \(k\), the inner sum in \(\ket{\phi_{|S\cap L|}}\) is an equal-amplitude superposition of all basis states of Hamming weight \(2k\), i.e., a Dicke state. On \(r\) qubits, the Dicke state of Hamming weight \(k\) is defined as
\begin{align}
    \ket{D_k^r} = \binom{r}{k}^{-1/2}\sum_{\substack{V \subseteq [r] \\ |V|=k}}\ket{V}.
\end{align}
Therefore, \(\ket{\phi_{|S\cap L|}}\) can be rewritten as
\begin{align}
    \ket{\phi_{|S\cap L|}} =\frac{1}{N_{S,L}}\sum_{k=0}^{\lfloor |S \cap L|/2\rfloor}\frac{1}{(2k-1)!!}\binom{|S\cap L|}{2k}^{1/2}\ket{D_{2k}^{|S \cap L|}}.
\end{align}

There exists a unitary operator \(\hat{U}_{r,k}\) with \(\hat{U}_{r,k}\ket{1^{l}0^{r-l}} = \ket{D_{l}^r}\) for all \(l \leq k\), which can be implemented with \(O(r^2)\) single- and two-qubit gates~\cite{bartschi2019deterministic}. Therefore,
\begin{align}
    \ket{\eta_{|S\cap L|}} := \hat{U}_{|S \cap L|,|S\cap L|}^\dagger \ket{\phi_{|S\cap L|}} = \frac{1}{N_{S,L}}\sum_{k=0}^{\lfloor |S \cap L|/2\rfloor}\frac{1}{(2k-1)!!}\binom{|S\cap L|}{2k}^{1/2}\ket{1^{2k}0^{|S\cap L|-2k}}.
\end{align}
 Let \(CR_y^{(i,j)}(\theta)\) denote the controlled-\(R_y\) gate with qubit \(i\) as control and qubit \(j\) as target, and let \(CR_y^{(0,j)}(\theta)\) denote the single qubit rotation \(R_y(\theta)\) on qubit \(j\). Then \(\ket{\eta_{|S\cap L|}}\) can be prepared from \(\ket{0^{|S\cap L|}}\) by the operator \(\hat{B}_{|S\cap L|}\), which consists of \(O(|S \cap L|)\) gates:
\begin{align}
    \hat{B}_{|S\cap L|}\ket{0^{|S \cap L|}} := \prod_{k=0}^{\lfloor|S\cap L|/2\rfloor - 1}CX^{(2k+1,2k+2)}CR_y^{(2k,2k+1)}(\theta_k)\ket{0^{|S \cap L|}} = \ket{\eta_{|S\cap L|}}
\end{align}
with
\begin{align}
    \theta_k = 2\arccos\left(\frac{\binom{|S \cap L| }{2k}^{1/2}/(2k-1)!!}{\sqrt{\sum_{l=k}^{\lfloor|S\cap L|/2\rfloor}\frac{1}{(2l-1)!!^2}\binom{|S\cap L|}{2l}}}\right).
\end{align}
Finally, let \(\hat{B}_{S,L}\) and \(\hat{U}_{S,L}\) denote \(\hat{B}_{|S\cap L|}\) and \(\hat{U}_{|S\cap L|,|S \cap L|}\) acting on the qubits \(S \cap L\) among the last \(p\) qubits, and define \(\hat{G}_{S,L} := \hat{B}_{S,L}
^\dagger \hat{U}_{S,L}^\dagger\). Then
\begin{align}
    \hat{G}_{S,L}\prod_{i \in L}CX^{(2n+i,i)}\ket{\psi_{S,L}} = \ket{S}\ket{0^{p}}.
\end{align}
Consequently, the loop-hafnian state \(\ket{\psi_{S,L}}\) is mapped to the basis state \(\ket{S}\ket{0^p}\) with \(O(p^2)\) gates.

\bibliography{apssamp}

\end{document}